\documentclass[aps,prl,twocolumn,showpacs,superscriptaddress,amsmath,amssymb]{revtex4}
\usepackage{epsfig}
\usepackage{graphicx}
\usepackage{dcolumn}
\usepackage{bm}
\usepackage{color}

\begin{document}

\title{Scanning Tunneling Microscopy in the superconductor LaSb$_2$}

\author{ \surname{J. A. Galvis}}
\author{ \surname{H. Suderow$^*$}}
\author{ \surname{S. Vieira}}

\affiliation{Laboratorio de Bajas Temperaturas, Departamento de F\'isica de la Materia
Condensada, Instituto de Ciencia de Materiales Nicol\'as Cabrera,
Facultad de Ciencias, Universidad Aut\'onoma de Madrid, E-28049 Madrid,
Spain}

\affiliation{Condensed Matter Physics Center (IFIMAC), Universidad Aut\'onoma de Madrid, E-28049 Madrid,
Spain}

\affiliation{Unidad Asociada de Bajas Temperaturas y Altos Campos Magn\'eticos, UAM/CSIC, Cantoblanco, E-28049 Madrid, Spain}

\author{ \surname{S. L. Bud'ko}}
\author{ \surname{P. C. Canfield}}
\affiliation{Ames Laboratory and Department of Physics and Astronomy, Iowa State University, Ames, IA50011, USA}

\date{\today}

\begin{abstract}

We present very low temperature (0.15 K) scanning tunneling microscopy and spectroscopy experiments in the layered superconductor LaSb$_2$. We obtain topographic microscopy images with surfaces showing hexagonal and square atomic size patterns, and observe in the tunneling conductance a superconducting gap. We find well defined quasiparticle peaks located at a bias voltage comparable to the weak coupling s-wave BCS expected gap value (0.17 meV). The amount of states at the Fermi level is however large and the curves are significantly broadened. We find T$_c$ of 1.2 K by following the tunneling conductance with temperature.
\end{abstract}


\maketitle

\section{Introduction}

\begin{figure}[!ht]
\begin{center}
\includegraphics[width=0.35\textwidth]{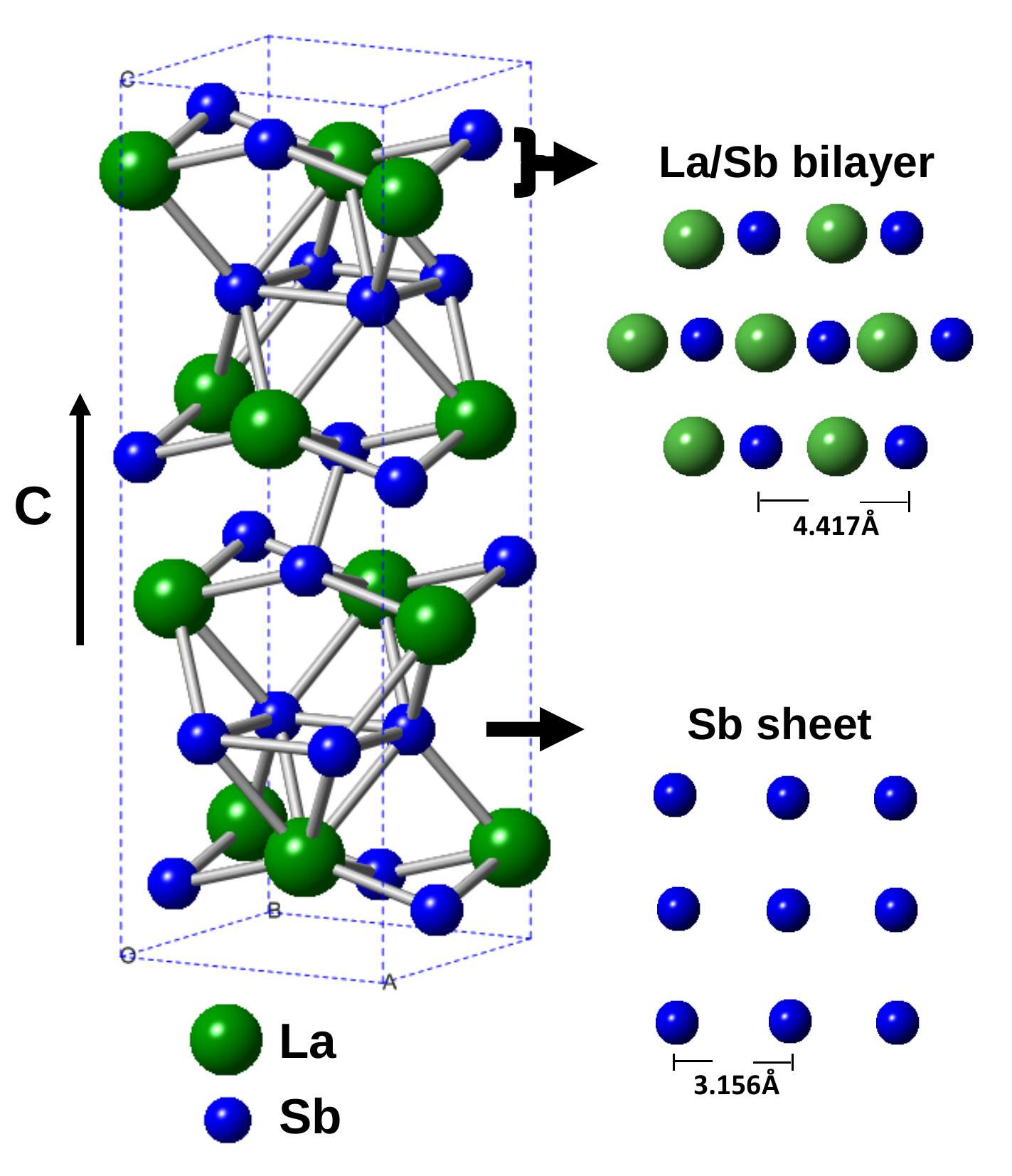}
\end{center}
\caption{Lattice structure of LaSb$_2$. A top view of one La/Sb bi-layers and one sheet of Sb is shown. Lattice parameters are $a=6.319{\AA}$, $b=6.1739{\AA}$ and $c=18.57{\AA}$.}
 \label{fig0}
\end{figure}

LaSb$_2$ is a layered rare-earth diantimonide which crystallizes in the layered orthorhombic SmSb$_2$ structure\cite{wang, canfield98, ditusa11}. La/Sb bi-layers of triangular prims alternate with two dimensional (2D) sheets of Sb as shown in Fig.\ref{fig0}. The isostructural rare-earth diantimonides RSb$_2$ (R=La-Nd, Sm) show highly anisotropic transport properties\cite{canfield98}. Resistivity measurements in LaSb$_2$ show that the temperature dependence is stronger for a current applied parallel to the c-axis than for a current applied in-plane\cite{canfield98,ditusa11}. On the other hand, susceptibility measurements show that LaSb$_2$ is a weakly temperature-dependent isotropic diamagnet\cite{canfield98,ditusa11}. In-plane magnetoresistance measurements at low temperatures (2 K) give an anisotropy for in-plane and c-axis applied magnetic fields that is very pronounced. Moreover, the in-plane magnetoresistance is linear and does not saturate up to 45 T.  LaSb$_2$ has been proposed as a good candidate to make high magnetic field sensors\cite{Young03}. The observed linear magnetoresistance is at odds with usual magnetic field dependence of the magnetoresistance\cite{Abrikosov03}. In metals, the magnetoresistance can saturate when Fermi surface is closed; remain quadratic up to high fields and then saturate when Fermi surface is open; or it can continously evolve eventually showing quantum oscillations in compensated metals with equal number of electrons and holes\cite{Alers53,Yan10}.

On the other hand, at higher temperatures, the magnetoresistance exhibits a pronounced maximum above 20 K, whose origin is yet unknown\cite{canfield98}. As there are no evidences for magnetic correlations in this material, the possible relationship to some sort of magnetic field induced charge density wave (CDW) has been highlighted in Ref.\cite{canfield98}. This possibility has been, in turn, used to try to explain the low temperature linear dependence of the magnetoresistance\cite{Young03,Goodrich04}. Yet another possible explanation is a linear energy spectrum\cite{Abrikosov99}. The compound LaAgSb$_2$, which also shows a large positive magnetoresistance up to 18 T\cite{Myers99,Wang12}, exhibits an anomaly in the resistance around 210 K which was associated to a CDW transition\cite{Song03,Myers99,Lue07}. It has been noted that other layered materials such as 2H-NbSe$_2$ and 2H-TaSe$_2$, where CDWs open at, respectively, 30 K and 120 K, also show roughly linear magnetoresistance\cite{Wilson75,Monceau2012}. In spite of all these hints, until now there is no evidence for CDW in LaSb$_2$, including most recent optical conductivity measurements\cite{ditusa11}.

The Fermi surface has been measured using de Haas-van Alphen oscillations in Ref.\cite{Goodrich04}. There are three Fermi surface sheets, two of them nearly two-dimensional (cylindrical along c), and one nearly spherical. All of them are very small, pointing out that LaSb$_2$ has a small number of carriers. Band structure calculations give further additional two bands, yet unobserved in de Haas van Alphen. Photoemission shows that Sb 5p states, hybridized with La 5d states, dominate the Fermi surface features, and that electronic properties are rather two-dimensional\cite{Acatrinei03}.

Among the rare earth diantimonides, LaSb$_2$ is unique in that it shows a superconducting state. Superconducting features are not clearly established in this material. Early reports give a very low T$_c$ of 0.4 K\cite{hullinger77}. However, more recent electrical transport and magnetic susceptibility measurements show a very broad transition, with a fragile onset as high as 2.5 K\cite{Guo11,ditusa11}. The resistance evolves then to roughly zero around 0.4 K-0.5 K. The superconducting properties exhibit strong dependence with the pressure, which reduces the anisotropy and sharpens the transition, with a T$_c=$ 2.1 K at 4 kbar\cite{Guo11}. Here we provide scanning tunneling microscopy and spectroscopy (STM/S) measurements in the superconducting phase of LaSb$_2$. We observe the atomic lattice at temperatures down to 150 mK.  We find a critical temperature of T$_c$=1.2 K, which is within the range of the broad transition found in macroscopic measurements\cite{ditusa11, Guo11}.

\section{EXPERIMENT}

We use a homebuilt STM/S system installed in a dilution refrigerator with an energy resolution in the tunneling spectroscopy of 0.15 K. This system is inserted in a vector magnetic field solenoid and has a similar construction to the one described in Ref.\cite{Suderow11}. We use a tip prepared from a gold wire cut with a clean blade, which is cleaned in-situ\cite{Rodrigo04}. Single crystals of LaSb$_2$ were grown from high-purity La and Sb in excess of Sb flux\cite{Canfield92,CanfieldEuroschool,Canfield91}. Large residual resistance ratios are found, of about 60, as in Ref.\cite{canfield98}. The crystals grow as plates with the c-axis perpendicular to the plates. They consist of soft Aluminium-foil like sheets that can be peeled off by glueing a stick and pushing it. We did not find good tunneling conditions in surfaces obtained immediately after an in-situ low-temperature cleave. After heating to room temperature, we realized that this was due to loose sheets remaining on the surface giving bad tunneling conditions. By cleaving the sample at ambient and removing those loose sheets using tweezers, we were able to obtain optically flat and shiny surfaces with no loose sheets. We measured two different samples, studying about four scanning windows in different regions of sample. Of course, surface contamination cannot be totally avoided when preparing the sample at ambient conditions. However we were able to find good scanning conditions at low temperatures, with reproducible imaging, being independent of the tunneling conductance. In the topographic images, the tip-sample bias voltage is fixed at 2 mV and the tunneling current remains constant while the tip is scanned over the sample surface at a tunneling conductance of order of or below a $\mu$S. The conductance vs bias voltage curves are obtained, as usual, by cutting the feedback loop and numerically differentiating the obtained current vs voltage curves\cite{Suderow11,Hermann01}. Tunneling conductance is normalized to one at bias voltages above one mV.

\section{RESULTS}

\begin{figure}[!ht]
\begin{center}
\includegraphics[width=0.44\textwidth]{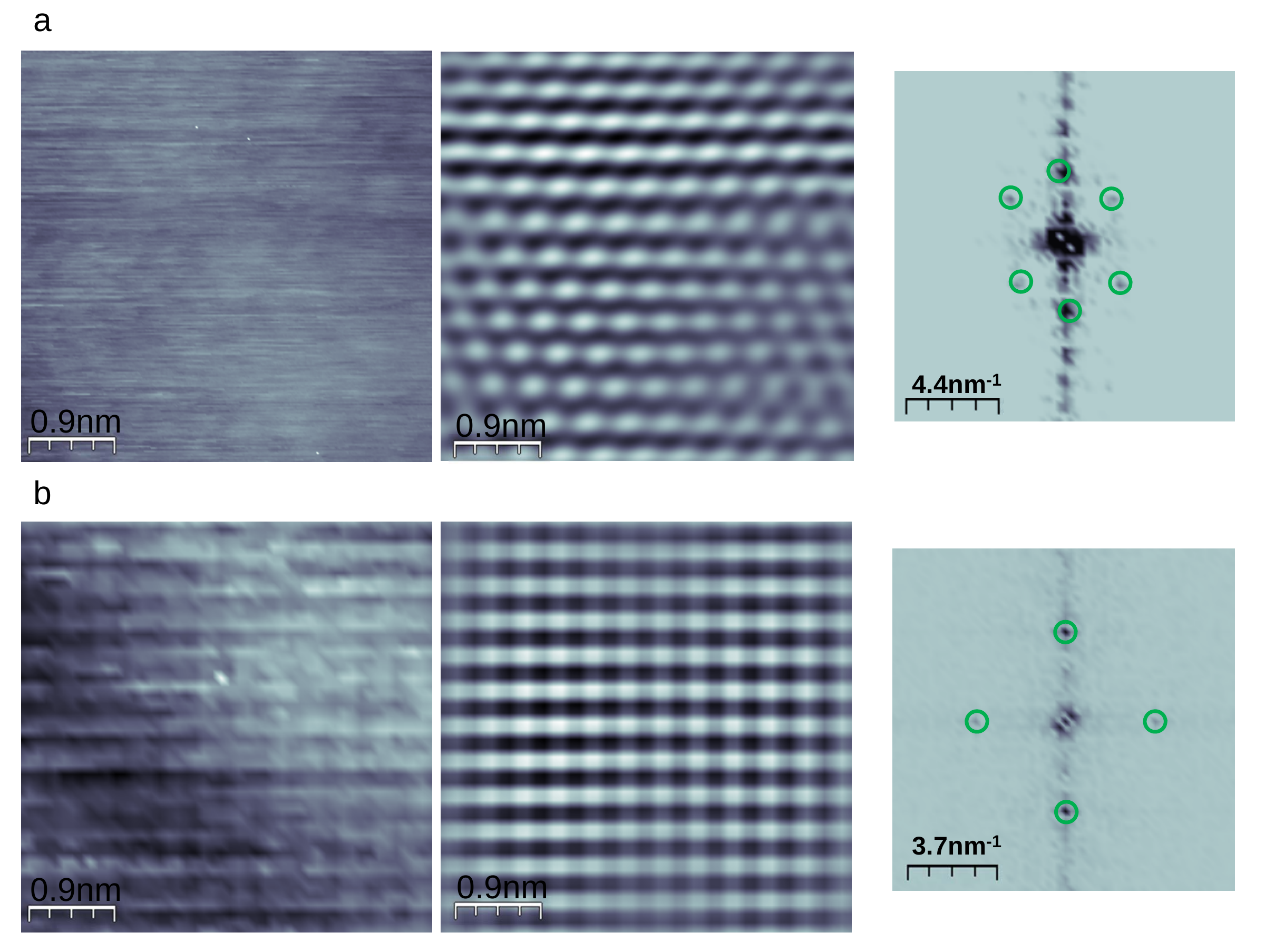}
\end{center}
\caption{In the left panels we show atomic resolution topography images at 0.15 K on two different LaSb$_2$ surfaces, with (a) hexagonal and (b) square atomic sized shapes. These images are unfiltered. The corresponding Fourier transforms is given in the right panels. Middle panels are Fourier filtered images, made by filtering out everything except green circles, which show the position of the Bragg peaks due to the atomic modulation.\protect\cite{Horcas07}}
 \label{fig1}
\end{figure}

In Fig.\ref{fig1} we show atomic resolution features in the topography images taken on different surfaces of the sample at 0.15 K. We can find hexagonal and also square atomic lattice arrangements (Fig.\ref{fig0}). The obtained geometry is in good agreement with the cuts on the structure shown in the right panels of Fig.\ref{fig0}. The lattice parameters also coincide with those measured with X-ray scattering\cite{Young03}. Accordingly, the surfaces showing square shapes are probably made out of Sb atoms and of La in those showing hexagonal shapes.

\begin{figure}[!ht]
\begin{center}
\includegraphics[width=0.4\textwidth]{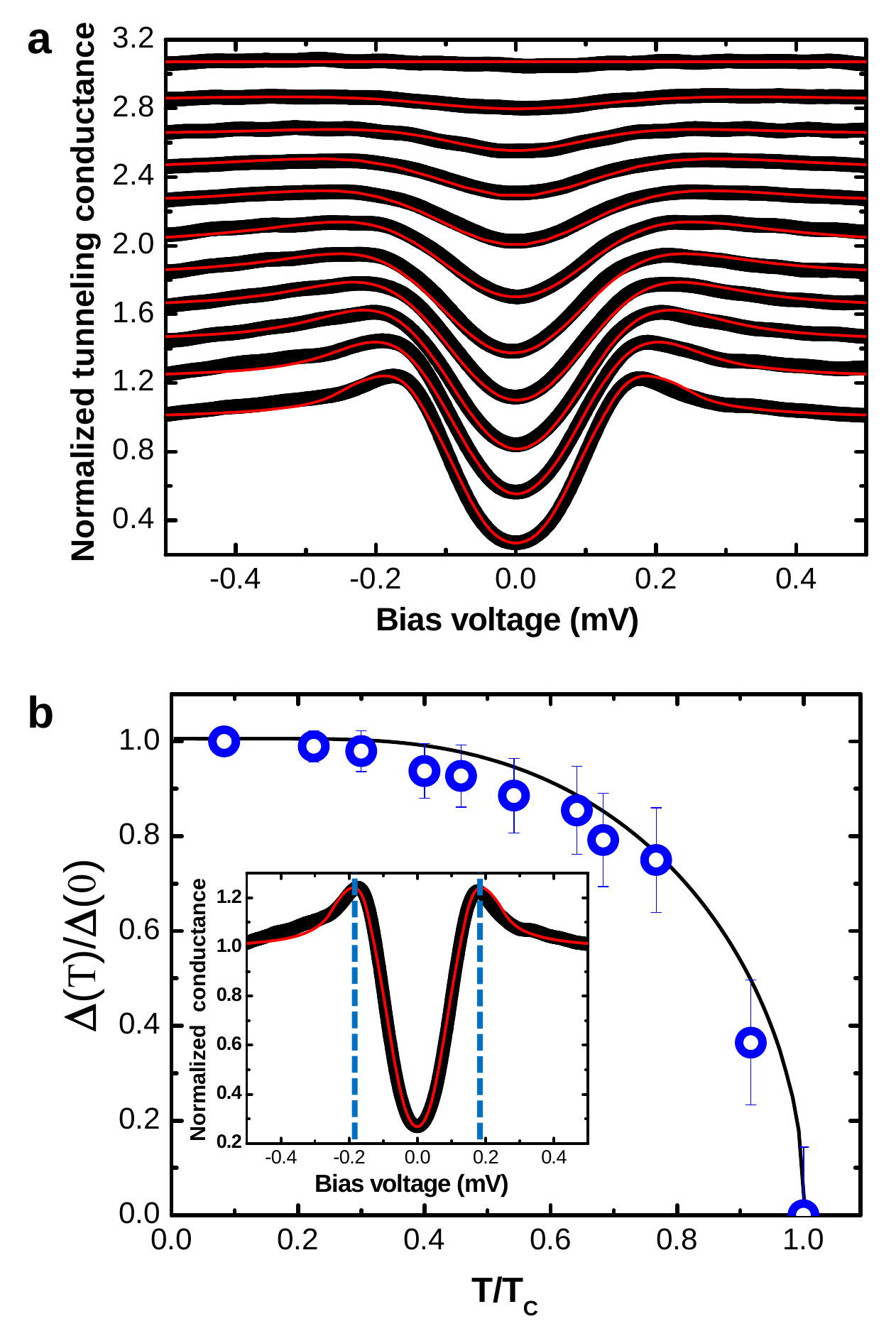}
\end{center}
\caption{a) Temperature dependence of the experimental tunneling conductance (black curves) and calculated conductance (red line). The data are, from bottom to top, taken at 0.15, 0.27, 0.36, 0.48, 0.55, 0.65, 0.77, 0.82, 0.92, 1.1 and 1.2 K. b) Temperature dependence of the position of the quasiparticle peak in the density of states used to calculate the red lines in a. The black solid line is the temperature dependence of the superconducting gap in BCS theory. Inset shows again the lowest temperature curve in a. Blue dashed lines give quasiparticle peak positions for a weak coupling BCS superconductor with T$_c$ of 1.2 K.}
\label{fig2}
\end{figure}

\begin{figure}[!ht]
\begin{center}
\includegraphics[width=0.42\textwidth]{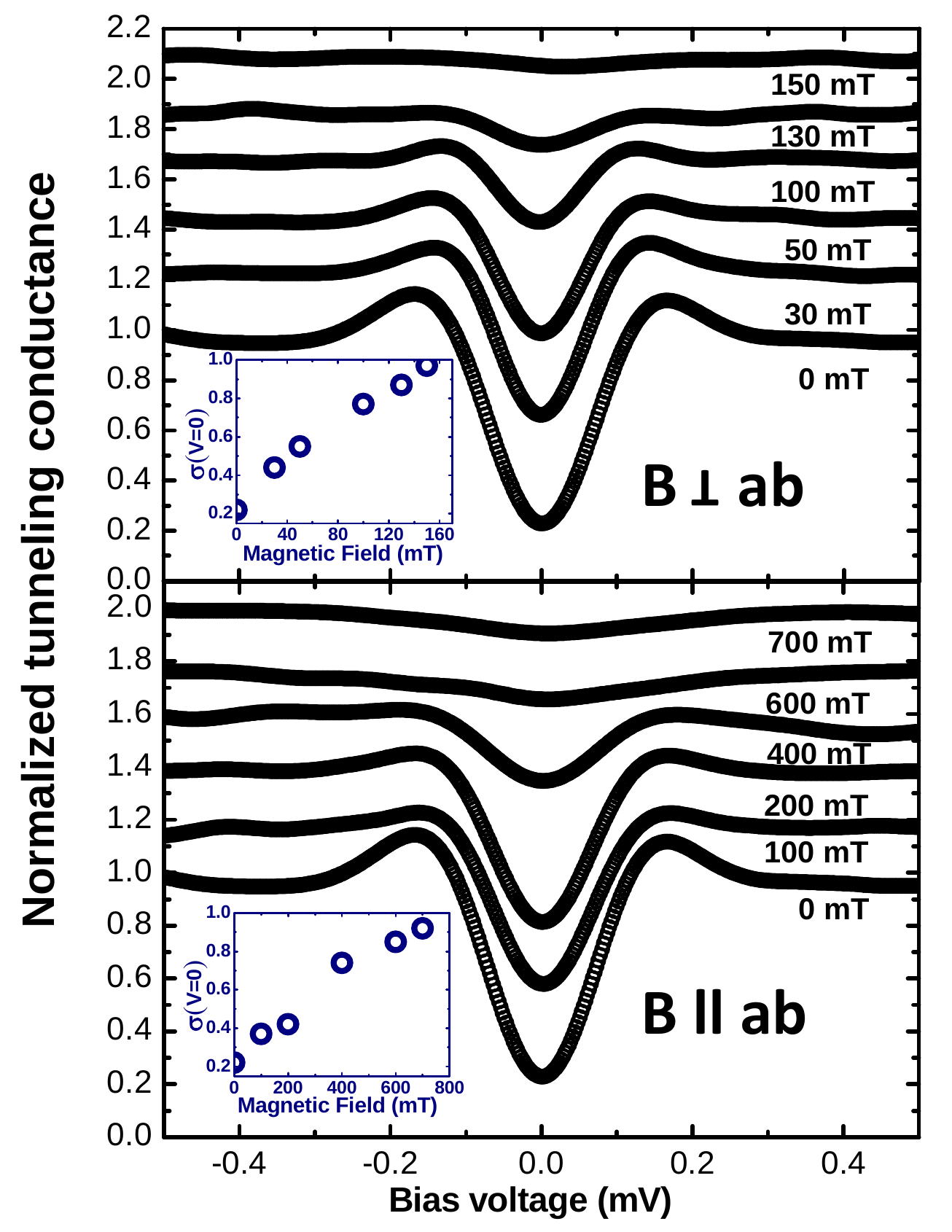}
\end{center}
\caption{Magnetic field dependence of tunneling conductance curves at 0.15 K for magnetic field applied perpendicular (a) and parallel (b) to the ab-plane. Insets show the magnetic field dependence of the zero bias conductance $\sigma(V=0mV)$.}
\label{fig3}
\end{figure}

Fig.\ref{fig2}a shows the tunneling conductance vs bias voltage and its temperature dependence. We find superconducting features with rounded quasi-particle peaks and a large conductance at zero and low bias. We could not observe clear atomic size variations of the superconducting features, as those previously measured in other compounds\cite{Guillamon08,Galvis12,Fischer07}. The superconducting features disappear fully into a flat tunneling conductance at about 1.2 K. At the lowest temperatures, the tunneling conductance curves show rounded quasi-particle peaks whose maximum is located at 0.17 mV. This can be fitted  (red line in inset of Fig.\ref{fig2}b) to a density of states showing a large amount of states at the Fermi level (25\% of the high bias voltage conductance) and a distribution of values of the superconducting gap, which ranges from 0.02 meV to 0.2 meV. The density of states and tunneling conductances are calculated by simply introducing a gap distribution in the BCS density of states formula, as made earlier in several superconductors showing multiband properties \cite{Rodrigo04b,Crespo06,Guillamon08}). The value for the superconducting gap $\Delta$ expected from the weak coupling equation $\Delta=1.76k_BT_c$ for $T_c=1.2K$, gives $\Delta=$0.182 meV (highlighted with the dotted blue line in Fig.\ref{fig2}b). When we increase the temperature, the tunneling conductance curves can be fitted by a density of states similar to the one at lowest temperatures, but with the gap distribution gradually shifted to lower energies when increasing temperature. The temperature dependence of the quasiparticle peak position in the corresponding density of states is shown in Fig.\ref{fig2}b, and is close to but somewhat below the BCS temperature dependence for the gap (black line in Fig.\ref{fig2}b).

The magnetic field dependence of the tunneling conductance is shown in top and bottom panels of Fig. \ref{fig3} with the field applied perpendicular and parallel to the plane. The curves are made following the same method as in Ref.\cite{Galvis12b}. We modify the scanning window where the curves are taken on the surface to find the largest value of the gap over the sample. The magnetic field leads to an additional broadening and flat curves above the critical field. We find an anisotropic critical field with a factor of approximately 4 between the parallel and perpendicular field. Using $H_{c2\bot}=\phi_0/2\pi \xi_\bot^2$ and $H_{c2\|}=\phi_0/2\pi \xi_\bot \xi_\|$ we find in-plane and c-axis coherence lengths, $\xi_\bot \sim$ 46 nm and $\xi_\| \sim$ 10 nm. Such a high value of in-plane coherence length $\xi_\bot$ requires rather large flat areas comprising several 100 nm to be able to see vortices. We did not obtain such large flat areas in our experiment.

\section{DISCUSSION AND CONCLUSIONS}

First, let us remark that the atomic resolution topography images and its associated Fourier transform (see Fig. \ref{fig1}) do not show any trace of modulations different to the atomic periodicity. Therefore, we find no evidence for CDW. This does not totally rule out its existence, but likely the CDW corresponds either to large wavelength modulations of the atomic lattice, above the size of our atomic images (roughly 10 lattice constants), or it occurs along the c-axis. The tunneling conductance up to 5 mV, which is the range we have explored, is also flat, showing that there is no pseudogap like behavior in this energy range. This is relevant, because the wide resistive transition could be related pre-formed pairs or some sort of pseudogap behavior as in the cuprates\cite{Fischer07}. The  features observed here at the surface of LaSb$_2$ are instead those of a good metal.

Regarding superconductivity, the shape of the tunneling conductance curves measured here present strong broadening. There is a distribution of values of the superconducting gap and a large amount of states close or at the Fermi level. The critical temperature is well defined and relatively high as compared to other measurements, $T_c=1.2$ K. In particular, it is higher than the low temperature part of the resistive transition, which starts at about 0.4 K\cite{hullinger77,Guo11}. On the other hand, upper critical magnetic fields obtained from resistivity are significantly smaller than the values reported here (0.02 T with the field applied along the c axis and 0.08 T with the field applied in plane, from resistivity\cite {Guo11}, and 0.15 T and 0.7 T found here). Superconducting features found in the STM tunneling conductance are more robust than in macroscopic resistance and susceptibility experiments. The observation of the atomic lattice points that the electronic properties of the surface show the intrinsic behavior of this material. 

STM probes superconductivity on the very few last layers, whereas the bulk properties are influenced by the in depth coupling among layers. Bulk superconductivity is stabilized by pressure, showing sharper superconducting transitions above 2 K\cite{Guo11}. This suggests that the large ambient pressure resistive transition could be related to layer decoupling and stacking faults in the bulk, which is reduced by applying pressure.

It is interesting to compare these results to experiments in layered transition metal dichalchogenide material 2H-TaSe$_2$, which shows an extremely small bulk critical temperature of 0.15 K\cite{Galvis12}. On compressing layers by applying pressure, T$_c$ increases rapidly to above 1 K with some kbar \cite{Smith75}. In 2H-TaSe$_2$, STM measurements show significantly increased critical temperature (up to 1 K) with respect to the bulk\cite{Galvis12}. At the same time, broadened gap features appear instead of the fully gapped superconductivity observed in similar materials such as 2H-NbSe$_2$\cite{Hermann08}.

In LaSb$_2$ the surface superconducting properties have higher T$_c$ than the resistive transition and we also observe broadened superconducting tunneling conductance features. The comparison with the dichalchogenides suggests that surface superconducting properties can be more robust and well defined than the bulk properties in layered materials. The last layers have possibly relaxed internal strains present in the bulk, eliminating inhomogeneous behavior observed in macroscopic properties.

We should note that, although the last layers show a single clean transition, the high amount of states close to the Fermi level points to anomalous superconducting properties. Its origin remains unclear, and they could be related to pair breaking effects due to disorder or coupling to nearby layers or regions with different superconducting properties. The broadened quasiparticle peaks show a wide distribution of values of the superconducting gap, mostly below the expected BCS value. This is compatible with multigap scenarios observed in two-band superconductors in layered materials\cite{Rodrigo04b,Guillamon08,Zehetmayer13,Liu01,Rubio01,Bascones01}. Multiband or multigap superconductivity can be expected within the Fermi surface features of LaSb$_2$\cite{Goodrich04}.

The appearance of stronger superconductivity at the surface suggests on the other hand that few or single layer systems could show superconducting properties. The fabrication of few layer sheets of LaSb$_2$ and similar compounds should be feasible, as they are easily peeled off and have a shape which is very similar to transition metal dichalchogenides. In the latter case, single layers have been synthesized already, showing in some cases superconducting correlations\cite{Galvis12,Castellanos10,Staley09,Ye12}. LaSb$_2$ belongs to the family of RSb$_2$, where R is a rare earth\cite{canfield98}. All compounds of the series appear in form of similarly shaped layered crystals. Having a rare earth in the crystal structure, they can lead to further interesting superconducting and magnetic properties not found in transition metal dichalchogenide layers.

In summary, we have observed the crystalline structure of LaSb$_2$ by means of atomic resolution STM measurements. The atomic lattice images at 0.15 K do not show signatures of charge density wave. In the tunnelling conductance curves, we observe broadened superconducting features disappearing at 1.2 K. The observed behavior can be related to the layered structure. We would like to highlight the remarkable combination of high purity samples, as seen in the residual resistance ratio, and superconducting features that are more robust and better defined at the surface.

This work was supported by the Spanish MINECO (Consolider Ingenio Molecular Nanoscience CSD2007-00010 program, FIS2011-23488, ACI-2009-0905), by the Comunidad de Madrid through program Nanobiomagnet and by COST MP1201. Work at the Ames Laboratory was supported by the US Department of Energy, Basic Energy Sciences, Division of Materials Sciences and Engineering under Contract No. DE-AC02-07CH11358.

$^*$ Corresponding author, hermann.suderow@uam.es.


\providecommand{\noopsort}[1]{}\providecommand{\singleletter}[1]{#1}%

\end{document}